\documentclass[twocolumn,preprintnumbers,showpacs,aps,prb]{revtex4}

\usepackage{graphicx,times,epsfig,color}

\usepackage{comment}

\begin{document}

\def\salto{\vskip 1cm}
\def\lgr{\langle\langle}
\def\rgr{\rangle\rangle}

\title{Lanczos transformation for quantum impurity problems in $d$-dimensional lattices: application to graphene nanoribbons}
%
\author{C. A. B\"usser}
\email[Corresponding author: ]{carlos.busser@gmail.com}
\affiliation{Department of Physics and Arnold Sommerfeld Center for Theoretical Physics, Ludwig-Maximilians-University Munich, Germany}
\author{G. B. Martins}
\affiliation{Department of Physics, Oakland University, Rochester, MI 48309, USA}
\affiliation{Materials Science and Technology Division, Oak Ridge National Laboratory, Oak Ridge, Tennessee 37831, USA}
\author{A. E. Feiguin}
\affiliation{Department of Physics, Northeastern University, Boston, Massachusetts 02115, USA}

\begin{abstract}
We present a completely unbiased and controlled numerical method to solve quantum impurity problems in {\it d}-dimensional lattices. 
This approach is based on a canonical transformation, of the Lanczos form, where the complete lattice Hamiltonian is exactly 
mapped onto an equivalent one dimensional system, in the same spirit as Wilson's numerical renormalization, and Haydock's recursion method. We introduce many-body interactions in the form of a Kondo or Anderson impurity and we solve the low-dimensional problem using the density matrix renormalization group. The technique is 
particularly suited to study systems that are inhomogeneous, and/or have a boundary.  
 The resulting dimensional reduction translates into a reduction of the scaling of 
the entanglement entropy by a factor $L^{d-1}$, where $L$ is the linear dimension of the original $d$-dimensional lattice. This allows one to calculate the ground state 
of a magnetic impurity attached to an $L\times L$ square lattice and an $L\times L \times L$ cubic lattice with $L$ up to 140 sites. 
We also study the localized edge states in graphene nanoribbons by attaching a magnetic impurity to the edge or the center of the system. 
For armchair metallic nanoribbons we find a slow decay of the spin correlations as a consequence of the delocalized metallic states. 
In the case of zigzag ribbons, the decay of the spin correlations depends on the position of the impurity. If the impurity is situated in the bulk 
of the ribbon, the decay is slow as in the metallic case. On the other hand, if the adatom is attached to the edge, the 
decay is fast, within few sites of the impurity, as a consequence of the localized edge states, and the short correlation length.
The mapping can be combined with ab-initio band structure calculations to model the system, and 
to understand correlation effects in quantum impurity problems starting from first principles.

\end{abstract}
\maketitle


\section{Introduction}
The Kondo problem describes a magnetic impurity embedded in a Fermi sea\cite{HewsonBook,cox1998exotic}. The spin of the impurity is screened by the spins of the electrons in the bulk forming a collective singlet state. Even though this is one of the most studied and better understood problems in condensed matter physics, we keep finding that our knowledge is incomplete, and open issues remain. In particular, there are questions that require a good description of the states and correlations in real space. For instance, great attention has been payed to the subtle issues of how to measure, detect, and characterize the so-called ``Kondo cloud''\cite{bergmann08,affleck2009kondo,FHM09,Busser10,Park2013}, or how to interpret the finite-size effects and competing energy scales arising when the conduction electrons are confined to a small spatial region in a ``Kondo box''\cite{schlottmann2001kondo,simon2002finite,simon2003kondo,kaul2006spectroscopy,hand2006spin}. Moreover, systems where the spatial position of the impurity is relevant (such as graphene nanoribbons, or topological insulators, where the physics of the edges is very different from the physics of the bulk) are quite non-trivial and represent a challenge to current state-of-the-art methods.

In the most general formulation of the problem the coupling of the impurity with the lattice has terms of the form $V_{k} d_{\sigma}^\dagger c_{k\sigma} \exp{(-i r_0 k)}$ (where $r_0$ is the position of the impurity and we used the usual notation, thus $d^\dagger_\sigma$ creates an electron at the impurity site with spin $\sigma$ and $c_{k\sigma}$ destroys an electron with spin $\sigma$ at the $k$ state). 
One of the usual approximations is to take simple forms for the dispersion $\epsilon_k$ or to ignore the momentum dependence of the couplings $V_k$, disregarding the information about the specific structure of the lattice\cite{nrg_wilson,andrei1983solution,tsvelik1983I,nrg_bulla}. Most commonly, one finds the impurity interacting with a wide band with a linear dispersion, via a contact local potential. This corresponds to the impurity being scattered only by $s$-wave states, and the problem can be mapped to an equivalent one-dimensional one.

 Systems where the effect of energy pseudogaps or Van-Hove singularities are important represent a challenge, and it is known, for instance, that the Kondo effect in graphene is quite non-trivial\cite{bulla1997anderson,uchoa2008localized,uchoa2011kondo,wehling2010orbitally,wehling2010theory,Lars}.
Moreover, approximations on the couplings $V_k$ make it difficult to deal with problems where the location of the impurity is relevant. As an example of this later situation we can mention graphene nanoribbons, or topological insulators, where the physics of the surface/edges is very different from the physics of the bulk. At the same time, experiments with adatoms on surfaces can be done with great degree of control using techniques such as STM \cite{crommie93,madhavan98,manoharam00,jamneala01,neel2008controlling,uchoa2009theory,trenes09,pruser2011long}, 
therefore, numerical methods that are able to deal with the spatial resolution of the Kondo problem are highly desirable. 

Numerical renormalization group (NRG), the optimal technique to study quantum impurities, does not work in a real space representation, but in energy space\cite{nrg_wilson,nrg_bulla}.  Until recently, quantum Monte Carlo (QMC) was the only computational technique that could offer some detail on the spatial structure of the correlations in dimensions larger than $d=1$ \cite{hirsch1986monte,gubernatis1987spin,werner2006continuous,gull2008continuous,gull2011continuous}.
Very recently, remarkable developments in the understanding of the NRG construction and its wave functions have offered a glimpse at the spatial structure of correlations around the impurity\cite{borda2007kondo,affleck2008friedel,mitchell2011real}. Moreover, using some ingenuity it is possible to introduce arbitrary dispersion/densities of states for the conduction electrons\cite{gonzalez1998renormalization,vzitko2009adaptive}. 

In this work, we introduce a new computational approach that enables one to work with the lattice in a real space representation at arbitrary dimensions with the density matrix renormalization group (DMRG) \cite{White1992,White1992a,White1993,Schollwock2005density}. 
The principal idea is to map, via a canonical transformation, the lattice onto a chain structure, in similar fashion to Wilson's original 
NRG formulation\cite{nrg_wilson}. In our scheme, same as in Haydock's recursion method\cite{Haydock1972,Haydock1975,Haydock1980,RecursionBook}, the information about the lattice structure, and the hybridization to the impurity, is 
completely preserved. Notice that this canonical transformation is exact, and the new representation of the Hamiltonian can 
be tackled with other methods especially suitable for solving the one-dimensional problem such as the embedded 
cluster approximation (ECA)\cite{Busser10}, or DMRG. 

The structure of the paper is as follows: In the next section we describe the technical aspects of the rotation to the Lanczos single particle basis that allows us to reduce the dimensionality of the problem. We also offer a geometrical interpretation of the transformation and the entanglement reduction. In Section III we explain how to study interacting quantum impurity problems on arbitrary lattice geometries using the DMRG method, illustrating with several proof of concept scenarios, such as 2d and 3d systems, and carbon nanotubes. Section IV focuses on the particular case of graphene nanoribbons and we numerically investigate the influence of the geometry and edge structure on the Kondo physics. We close with a summary and conclusions in Section V.

\begin{figure}
\vspace{0.25cm}
\epsfxsize=7.5cm \centerline{\epsfbox{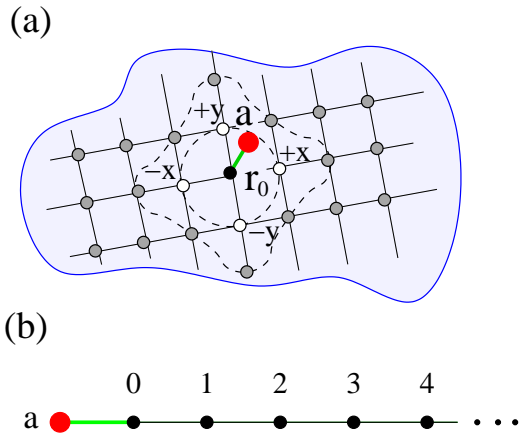}}
\caption{{\it Mapping of the system into a semi-chain}. 
Using the Lanczos method, a real lattice can be mapped into a 1d semi-chain.
The coupling between the impurity and the lattice, shown in panel (a), determines the construction of the 
Lanczos seed $|\Psi_0\rangle$. In this case, the seed consists of just a single orbital, which becomes 
the first site of the mapped chain, as shown in panel (b). The dashed lines connect the sites that form the Lanczos orbitals of the resulting 1d system.}
\label{figure1}
\vspace{0.25cm}
\end{figure}
\section{Mapping bands onto chains}

\subsection{Change of basis and Lanczos orbitals}

Let us start with a simple Hamiltonian with one impurity connected to one single lattice site. A more general case will be discussed below. The Hamiltonian is
\begin{equation}
H =  H_{\rm imp} + H_{\rm band} + V,
\label{hamiltonian}
\end{equation}
where $H_{\rm imp}$ is the many body impurity Hamiltonian, $H_{\rm band}$ is the lattice Hamiltonian and $V$ describes the impurity-lattice interaction.  
For the Anderson model, the impurity is described by 
\begin{equation}
H_{\rm imp} = V_g \sum_\sigma n_{d\sigma} + U n_{d\uparrow} n_{d\downarrow},
\label{Anderson_model}
\end{equation}
where $n_{d\sigma}$ is the occupation operator for an impurity with spin $\sigma$, $V_g$ is a gate potential, and $U$ 
parametrizes the on-site Coulomb repulsion between opposite-spin electrons.

The corresponding hybridization term is introduced as
\begin{equation}
V = -t' \sum_{\sigma} \left(d^\dagger_{\sigma} c_{{\rm r_0} \sigma} + {\rm h.c.}\right), 
\label{hybridization}
\end{equation}
where $d^\dagger_{\sigma}$ creates an electron at the impurity and $c_{{\rm r_0} \sigma}$ destroys an electron at position $\rm r_0$, 
the lattice site connected to the impurity.

Alternatively, one could introduce the impurity via a Kondo Hamiltonian, with an interaction
\begin{equation}
V = J_K \vec{S}_d \cdot \vec{S}_{\rm r_0}.
\label{Kondo}
\end{equation}

The main point of this proposal is to map a complex band structure, for example, a carbon nanotube, a graphene ribbon, or any 2d or 3d system, onto a semi-chain.  
This mapping can be done in such a  way that an impurity connected to one or more sites of a real lattice is, after the mapping, 
connected to a single site of a chain. The resulting one-dimensional problem can then be solved with a method of choice, such as ECA or DMRG.

In order to perform the mapping, we need to assume that the band describes non-interacting electrons via a quadratic tight-binding Hamiltonian.
 One can always map a complicated one-body Hamiltonian onto a semi-chain by applying the Lanczos method in the same spirit as 
Wilson's NRG or Haydock's recursion method\cite{Haydock1972,Haydock1975,Haydock1980,RecursionBook}. To start the change of basis, we choose the orbital that is connected to the impurity as seed for the iterative procedure\cite{Dagotto1994}(we drop the spin index for now):
\begin{equation}
|\Psi_0 \rangle = c^\dagger_{{\rm r_0}} |0\rangle.
\label{seed}
\end{equation}
The so-called Lanczos basis is constructed as:
\begin{eqnarray}
|\Psi_1\rangle &=& H_{\mbox{\rm band}} |\Psi_0\rangle - a_0~|\Psi_0\rangle \\
|\Psi_{i+1}\rangle &=& H_{\mbox{\rm band}} |\Psi_i\rangle - a_i |\Psi_i\rangle - b_i^2 |\Psi_{i-1}\rangle.
\label{recursion}
\end{eqnarray}
where,
\begin{equation}
a_i = \frac{\langle\Psi_i|H_{\mbox{\rm band}}|\Psi_i\rangle}{\langle\Psi_i|\Psi_i\rangle},
~~~~~~ b_i^2 = \frac{\langle\Psi_i|\Psi_{i}\rangle}{\langle\Psi_{i-1}|\Psi_{i-1}\rangle}.
\end{equation}
This allows one to construct a basis where the lattice Hamiltonian $H_{\rm band}$ is tri-diagonal: 
\begin{equation}\bigskip
H_{\mbox{\rm band}} = \pmatrix{a_{ 0}&b_{ 1}&0&0&\cdots\cr 
b_{1}&a_{1}&b_{2}&0&\cdots\cr 
0&b_{2}&a_{2}&b_{3}&\cdots\cr 
0&0&b_{3}&a_{3}&\cdots\cr 
\,\vdots\hfill&\,\vdots\hfill&\,\vdots\hfill&\,\vdots\hfill&&\cr}.
\label{Hband}
\end{equation}
This is an exact canonical transformation, from a single particle basis, onto another complete orthogonal basis. 

In Fig.\ref{figure1}(a) we show a schematic representation of the problem in real space. Panel (b) shows the equivalent Hamiltonian in the Lanczos basis. 
The off-diagonal matrix elements ($b_i,~i=1,2,...$) become effective hopping terms, while the diagonal ones ($a_i,~i=0,1,2,...$) introduce local 
chemical potentials. In second quantization, it reads:
\begin{equation}
H_{\rm band} = \sum_{i=0}^{N-2} a_i \tilde{n}_i + \sum_{i=0}^{N-3} b_{i+1} \left(\tilde{c}_{i\sigma}^\dagger \tilde{c}_{i+1,\sigma} + {\rm h.c.}\right),
\end{equation}
where the new ``tilde'' operators refer to the Lanczos orbitals along the chain and obey fermionic anti-commutation rules. Notice that the first orbital, corresponding to the one connected to the impurity and used as seed, remains unaltered after the transformation, {\it i.e}, $\tilde{c}_0 \equiv c_0$. As a consequence, the hybridization Hamiltonian $V$ remains unchanged.

In the process of generating the new basis we find that: 
\begin{enumerate}
\item $\langle\Psi_j|\Psi_i\rangle= \delta_{i,j} $

\item $\langle\Psi_j|H_{\mbox{\rm band}}|\Psi_i\rangle= b_j \delta_{j,i+1} + a_i \delta_{i,j} + b_i \delta_{i,j+1}$

\item The basis $\{~|\Psi_n\rangle\}$ is orthogonal, but it is not normalized.

\item Obviously, once one of the $b_n$ vanishes, the rest of the band will decouple.
In fact, all $|\Psi_m\rangle$, for $m>n$, will not be defined (the rest of the basis belongs to different channels/symmetry sectors). 

\item As $\langle\Psi_n|\Psi_n\rangle$ grows with $n$, numerical errors due to finite numerical precision might appear.
To avoid them, we have to re-normalize the states $|\Psi_{i-1}\rangle$ at each iteration.
\label{num1}

\item Another numerical error to consider is the loss of orthogonality between 
vectors for large $n$. To avoid that, and keep the basis complete, we orthogonalize 
each new vector with all the previous vectors at each iteration.
\label{num2}

\end{enumerate}

The transformation can also be applied to problems where either the band, or the impurity, have a more complex structure, as described in the following sections.

\subsection{Geometrical interpretation and entanglement}

The structure of the Lanczos orbitals can be interpreted geometrically in a very simple way, for a system such as the 
square lattice. The reference site where the impurity is attached becomes a center of symmetry for the 
point group operations of the lattice, as shown in Fig.\ref{figure1}. Since the hopping Hamiltonian obeys the same symmetries, 
the Lanczos orbitals will belong to the same symmetry sector as the seed orbital. If the impurity is attached 
locally to a single site of the lattice, and this site is chosen as seed, then all the Lanczos orbitals will have ``$s$-wave'' symmetry, invariant under rotations 
of the lattice. Strictly speaking, since we are not in the continuum, this corresponds to the trivial $\{x^2+y^2\}$ representation of the group or rotations $C_4$ of the square lattice.
All other {\it channels}, corresponding to different symmetry classes, will form their own independent chains, and will be completely 
decoupled from the impurity. This yields a remarkable result: in order to study a system with $L^d$ sites 
(where $L$ is the linear size, and $d$ the dimensionality of the problem), we only need to keep $N\sim {\cal O}(L)$ orbitals!

To clarify these ideas let us illustrate with a simple example: we focus on the nearest neighbors around the center of symmetry, and we label the corresponding single particle orbitals as $|\pm x\rangle$, $|\pm y\rangle$, as shown in Fig.\ref{figure1}(a). The first Lanczos orbital is simply given by 
\[
|\Psi_{1,\{x^2+y^2\}}\rangle = \frac{1}{2}\left[ |+x\rangle + |-x\rangle +|+y\rangle + |-y\rangle \right].
\]
There are three other single particle orbitals that are not coupled by the Hamiltonian. We could define them as, for instance:
\begin{eqnarray}
|\Psi_{1,\{x^2-y^2\}}\rangle & = & \frac{1}{2}\left[ |+x\rangle + |-x\rangle -|+y\rangle - |-y\rangle \right], \nonumber \\
|\Psi_{1,\{x\}}\rangle & = & \frac{1}{\sqrt{2}}\left[ |+x\rangle - |-x\rangle \right], \\
|\Psi_{1,\{y\}}\rangle & = & \frac{1}{\sqrt{2}}\left[ |+y\rangle - |-y\rangle \right], \nonumber 
\end{eqnarray}
where we have used the fact that the wavefunctions transform as the different irreducible representations of the point group. If we use these orbitals as seeds for the Lanczos iterations, each will generate another chain, or channel, but they will be totally decoupled from each other, and from the original center of symmetry, since the Hamiltonian will not mix different symmetry sectors. We can also see that as the distance from the center increases, the number of linearly independent wavefunctions that we can construct, and consequently, the number of channels will also increase.

\begin{figure}
\epsfxsize=7cm \centerline{\epsfbox{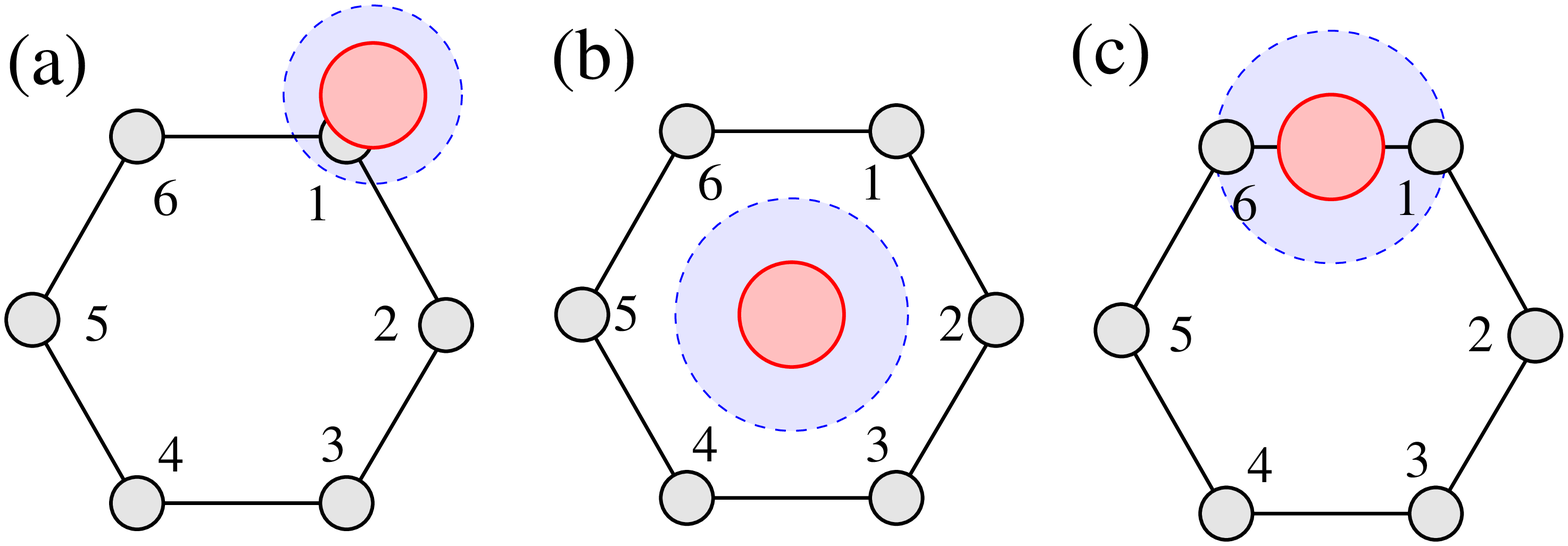}}
\caption{{\it Choosing the seed.} Examples of how the coupling between the impurity and the lattice will determine the construction of 
the seed. We are using a honeycomb carbon lattice, but the idea can be extended to any geometry.
(a) In the simplest case (and the one used for the calculations in the rest of the paper), the impurity is sitting on top of just one atom. 
(b) The impurity is connected to all six sites. 
(c) The impurity is on the bond between two C atoms.}  
\label{figure2}
\end{figure}

The orbitals are then defined by their symmetry sector, or channel, and their radial distance from the center of symmetry, which is 
correlated to the linear distance along the equivalent 1d chain. Notice that in a bipartite lattice, odd and even sites along 
the chain will correspond to orbitals with support on different sublattices, a result that will become important in the interpretation of numerical results.

As a consequence of these observations, we obtain an intuitive understanding of the behavior of the entanglement: The 
entanglement {\it per channel} between the region enclosed by an area of ``radius'' $L$ in the $d$-dimensional problem 
is exactly the same as the entanglement between the first $L$ sites of the chain, and the rest. In cases where the system under consideration is gapless, the 
chain is a critical one-dimensional system (for square and cubic lattices, for instance), and the von Neumann 
entanglement entropy is proportional to $\log(L)$ \cite{calabrese2004,calabrese2006}. All channels contribute to the 
entropy with similar factors. It can be seen that the number of channels is proportional to the area of the boundary $\sim L^{d-1}$. This yields a final result proportional to $L^{d-1} \log(L)$.
Therefore, based on these simple arguments, we can easily understand why free fermions in higher-dimensions have 
logarithmic corrections to the area law\cite{wolf2006,gioev2006,li2006,barthel2006}.
The advantage of our approach is that we only need to solve the problem in the channel 
that is coupled to the impurity, reducing the entanglement by a factor of $L^{d-1}$!

Notice that these ideas are basically a generalization of the one dimensional case studied in Ref.~\onlinecite{feiguin2011reducing}. In 
a one dimensional impurity problem, one can apply reflection symmetry and make a ``folding'' transformation, 
mapping the single particle orbitals onto bonding and antibonding states. In that case, the impurity couples only 
to the bonding channel, while the anti-bonding remains decoupled. This translates into a reduction of the entanglement by a factor of $2$.

We point out that in generic situations, the coupling between the impurity and the lattice could be along a different channel 
in a different symmetry sector. Moreover, it could couple to more than one channel at the same time. Tackling this problem is explained in the next section.

\subsection{Constructing the seed orbital}

\begin{figure}
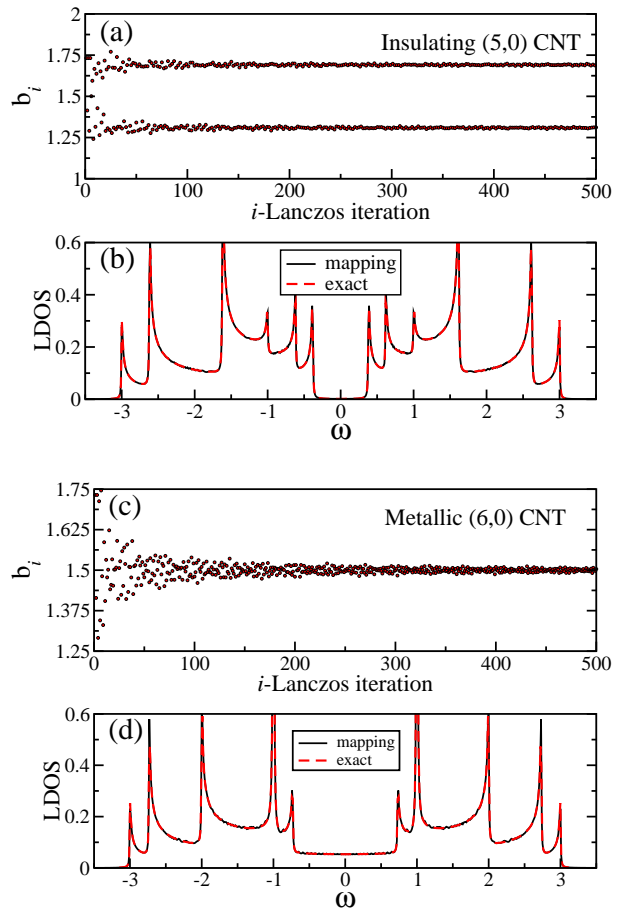

\epsfxsize=8cm \centerline{\epsfbox{figure3a-II.eps}}

\vspace{0.5cm}
\epsfxsize=8cm \centerline{\epsfbox{figure3b-II.eps}}
\caption{{\it Local density of states for CNT and hopping parameters for a single chain mapping.} 
Chain hopping vs. $n$ [panels (a) and (c)] and LDOS as function of $\omega$ [panels (b) and (d)] 
are shown for an insulating (top 2 panels) and a metallic (bottom 2 panels) CNT. 
For the LDOS, the exact solution was calculated using Green's functions [(red) dashed line]. 
The agreement with the mapping calculation [(black) solid line] is excellent.}
\label{figure3}
\end{figure}  

As discussed above, one of the central issues in the canonical transformation is how to pick the right seed $|\Psi_0\rangle$. 
This is determined by the chemistry of the problem, and therefore several situations may arise. 
In the simplest scenario, the impurity may be coupled to a single site, 
but, in a more general case, it may couple to many lattice sites. In Fig.~\ref{figure2}, as an 
illustration, we show the case of one adatom in a honeycomb lattice. The adatom can sit on top of one of the atoms 
of the lattice [panel (a)], or at the center of an hexagon [panels (b) and (c)]. 
In this latter case, depending on the hybridization with the lattice, the impurity may couple symmetrically 
to all six sites, as shown in panel (b), or may be connected along the bond between two carbon atoms, as illustrated in panel (c). 
The general expression for the impurity-lattice coupling Hamiltonian is given by 
\begin{equation}
V = -\sum_{i=1 \ldots 6, ~\sigma} \left( t_{{\rm 0} i} ~d_{\sigma}^\dagger ~c_{r_i\sigma} +~\mbox{h.c.}\right).
\end{equation}
 Then, we define a renormalized orbital
\begin{equation}
c_{r_0\sigma} = -\sum_i \frac{t_{{\rm 0} i}}{\tilde{t}_0} ~c_{r_i\sigma},
\end{equation}
and rewrite the coupling Hamiltonian as:
\begin{equation}
V = -\sum_{\sigma}  \left( \tilde{t}_0~d_{\sigma}^\dagger ~c_{r_0\sigma} +~\mbox{h.c.} \right),
\end{equation}
where we have used the definition $\tilde{t}_0=\sqrt{\sum_i t_{{\rm 0} i}^2}$.
Now, the seed can be defined as
\begin{equation}
|\Psi_0 \rangle = c^\dagger_{r_0} |0\rangle.
\end{equation}

Although here we illustrate the method with the example of an adatom, it could be generalized to the case of a substitutional impurity, or defect.
We have not considered the situation in which the impurity has more than one orbital. In such a case, the resulting problem would map on an impurity coupled to two chains. This problem will be studied in future work.

Notice that this is possible because the coupling Hamiltonian has only one-body terms, such as is the case of Anderson-type Hamiltonians. 
For coupling Hamiltonians containing many-body terms\cite{Jacob10,Lars}, if the impurity is connected 
to more than one site, it is not possible to choose such a simple seed.
In that case, the proper way to identify the coupling Hamiltonian is by doing a Schrieffer-Wolff transformation. 
The seed to use in that case can be obtained through a generalization of the ideas above.
%

\begin{figure}
\epsfxsize=7.5cm \centerline{\epsfbox{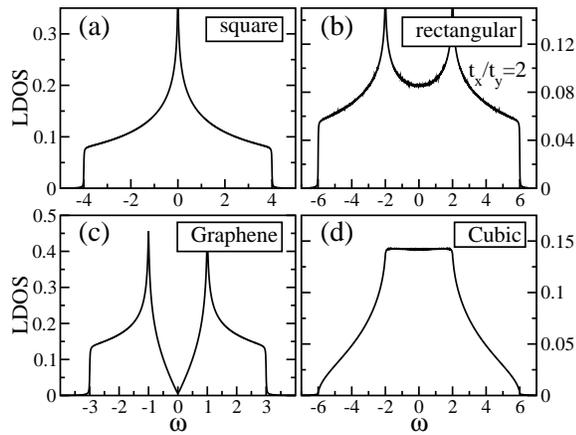}}
\caption{{\it LDOS for several lattices.} The LDOS is plotted as a function of $\omega$ for: (a) 2d square lattice with $800^2$ sites, (b) rectangular lattice with $800^2$ sites, 
(c) graphene sheet with $800^2$ sites, (d) 3d cubic lattice with $200^3$ sites.}
\label{figure4}
\end{figure}  
\subsection{Non-interacting results}

As a first benchmark test we present calculations of one-body properties for a system without impurities. This is basically a startightforward application of the recursion method for non-interacting tight-binding bands\cite{Haydock1972,Haydock1975}. Since this is a single body problem, results are known. 
We compare the exact local density of states (LDOS) at the position of the seed orbital, which, as mentioned above, remains unchanged by the basis transformation, 
with the ones obtained for the resulting chain system (see Fig.~\ref{figure1}). 

We start by performing the mapping for carbon nanotubes (CNTs) \cite{charlier2007electronic}. 
We point out that the recursion method has been already been used in this context \cite{triozon2004electrical}, and also for graphene \cite{lherbier2008transport,cresti2008charge}.
As it is well known, depending on its chirality, a CNT can be metallic or insulator\cite{nanotubes}.
In Fig.~\ref{figure3} we show results for the hopping parameters in the mapped chain [panels (a) and (c)] and the LDOS [(b) and (d)]. 
We analyze two zigzag CNTs: $\left(5,0\right)$, which is insulating (upper panels), and $\left(6,0\right)$, which is metallic (lower panels). 

In panels (a) and (b) we present the insulating case $\left(5,0\right)$, whose LDOS presents a gap separating the valence 
and conduction bands. Twelve Van Hove singularities can be observed, 
indicating a complex band structure. We compare the mapped results with those obtained by direct diagonalization [(red) dashed lines]. No difference between them can be observed. 
%

\begin{figure}
\epsfxsize=7.5cm \centerline{\epsfbox{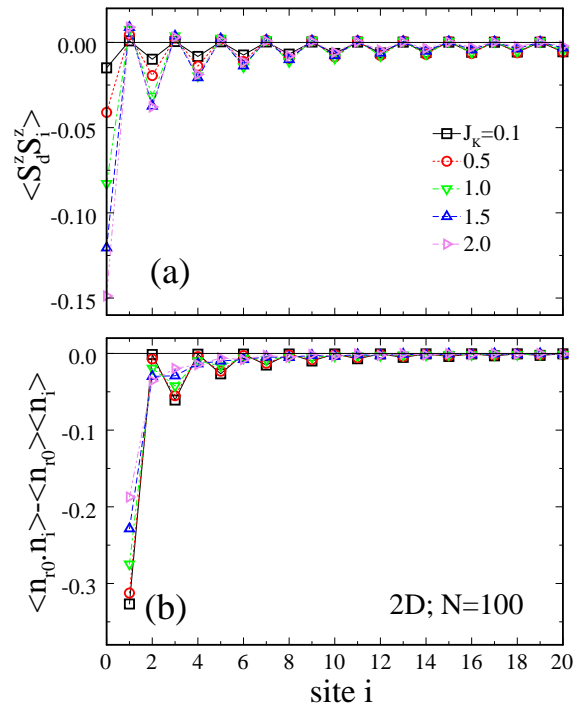}}
\caption{{\it Correlations for a 2d square lattice.} (a) Spin-spin correlations between a Kondo impurity and the 
rest of the chain, for different values of $J_K$, keeping $N=100$ sites. We show results for the first 20 sites of the chain. 
(b) density-density correlations between the first site of the chain, and the rest. All simulations were done at half-filling. 
{Notice that this system length corresponds to an actual square lattice with $L\sim140^2$ sites.}}
\label{corr_2d}
\end{figure}  

The hopping parameters for the resulting chain are shown in panel (a), and we observe two families of hoppings/energy scales, 
around $t_a\sim1.7$ and $t_b\sim1.3$. This repeated structure (A-B-A-B-A-B-A-B...) resembles a superlattice with a two-site basis (A-B). 
As a consequence, the band structure develops a gap. Of course, there are other lower frequency structures in the hopping elements associated to 
the other features of the spectrum. 

On panels (c) and (d) in Fig.~\ref{figure3} we show results for the $\left(6,0\right)$ metallic CNT. In this case we can see ten Van Hove 
singularities in the LDOS. Again, we compare with the exact solution and the agreement is excellent. 
As in this case there is no gap, the corresponding chain hopping parameters [panel (c)] do not present the dimer structure shown for the gaped LDOS [panel (a)]. 

\begin{figure}
\epsfxsize=7.5cm \centerline{\epsfbox{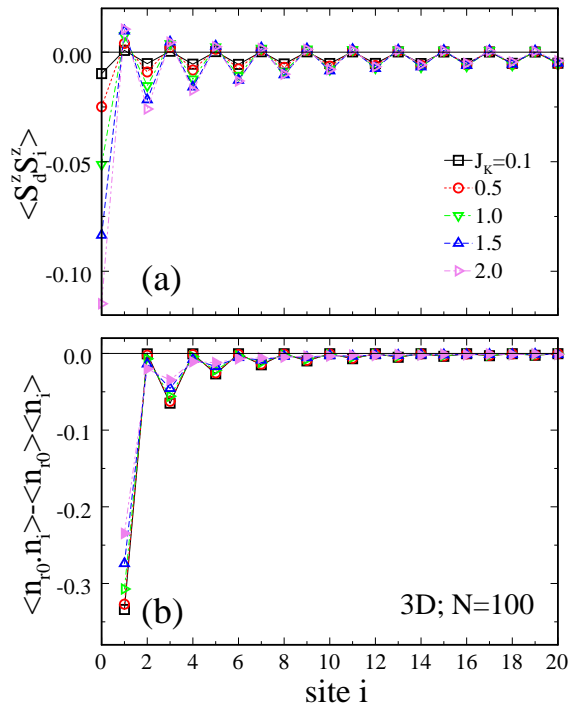}}
\caption{{\it Correlations for a 3d cubic lattice.} (a) Spin-spin correlations between a Kondo impurity and 
the rest of the chain, for different values of $J_K$, keeping $N=100$ sites. As in Fig.~\ref{corr_2d}, we show 
results for the first 20 sites of the chain. (b) density-density correlations between the first site of the chain, and the 
rest. Notice that this system size corresponds to an actual cubic lattice with a diagonal size of $100$ sites.}
\label{corr_3d}
\end{figure}  
%
As an illustration, in Fig.~\ref{corr_3d}, we also present results for higher-dimensional systems. As expected, we reproduce exactly all the features of 
the spectrum, such as the Van Hove singularity at $\omega=0$ for the 2d square lattice [panel (a)]. In 
the case of a rectangular lattice [panel (b)], where the lattice parameter in one direction is twice that in the other direction, we see that the Van Hove 
singularity splits into two peaks. 
We can also reproduce, after the mapping, the band structure of graphene [panel (c)], where we recover the massless Dirac dispersion at $\omega=0$. 
The 3d LDOS in panel (d) (for a cubic lattice) provides a simple test bed to study problems with a quasi-flat band, 
such as the one used in analytical treatments of the Kondo model, and its Bethe Ansatz solution.
Needless to say, the fact that all these well known results are here neatly reproduced is not surprising, as the proposed mapping is an exact canonical transformation.

Notice that the Lanczos orbitals will have suppport on a subspace and will not span the entire Hilbert space. This means that we are actually reproducing portion of the original energy spectrum and DOS. It is important to point out that the energy level corresponding to the Fermi energy can be determined by fixing the density of the chain. The other channels will also be partially filled, but they are not relevant to our calculation.


\begin{figure}
\epsfxsize=6.cm \centerline{\epsfbox{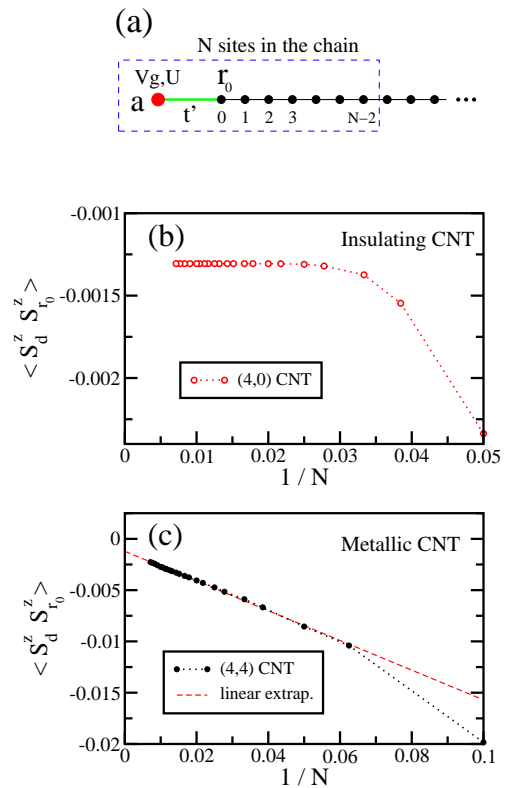}}

\vspace{0.1cm}
\epsfxsize=6.5cm \centerline{\epsfbox{figure7bc.eps}}
\caption{{\it Scaling of correlations in CNTs using DMRG.} (a) In order to extrapolate to the thermodynamic 
limit we first map the band Hamiltonian onto its tridiagonal Lanczos form. We study the convergence as a function 
of $1/N$, where $N$ is the number of sites kept in the actual simulation. 
Panels (b) and (c) show the spin-spin correlations as function $1/N$ for an adatom at the surface of a CNT. 
Panel (b) corresponds to an insulating $\left(0,4\right)$ CNT. In this case, as there are no extended states in the CNT, convergence is fast.
Panel (c) shows results for a metallic $\left(4,4\right)$ CNT. Due to the gapless metallic extended states, the convergence is slower. A linear 
extrapolation in $1/N$ is also shown in the figure [(red) dashed curve]. The parameters used are $U=1$, $V_g=-U/2$ and $t'=\sqrt{0.05}$.
}
\label{figure7}
\end{figure}  

\section{DMRG solution of the interacting problem}
\label{SecIV}

In order to study the many-body problem with an interacting impurity, we use the DMRG method, which is specially suitable for 
one-dimensional problems. We perform the mapping of the band Hamiltonian and solve the full interacting problem, with the impurity connected to 
the first orbital of the chain, which, as mentioned above, was used as the seed to construct the Lanczos basis. Obtaining the mapping for large 
systems is straightforward, but in order to do a DMRG calculation we limit the resulting chain sizes to a few hundred sites, and perform a finite size 
scaling in $1/N$, where $N$ is the linear dimension of the problem, as displayed in Fig.\ref{figure7}{a}. In all calculations we work at half-filling, keeping the DMRG truncation error below $10^{-9}$.

As a proof of concept, we first solve the Kondo Hamiltonian Eq.~(\ref{Kondo}) for a 2d square lattice. In Fig.~\ref{corr_2d}(a) we show the spin-spin correlations 
between the impurity and the Lanczos orbitals of the chain $\langle S^z_dS^z_i \rangle$ as a function of the distance from the impurity, 
for $0.1 \leq J_K \leq 2.0$. We see a characteristic power-law behavior, with alternating antiferromagnetic (AFM) and ferromagnetic (FM) signs. 
The ferromagnetic 
correlations within the same sublattice, for small values of $i$, clearly increase with $J_K$. The density-density correlations are plotted in panel (b). In 
this case, we show the correlations between the first site of the chain and the 
rest $\langle n_{\rm r_0}n_i \rangle-\langle n_{\rm r_0}\rangle\langle n_i\rangle$. In Fig.~\ref{corr_3d} we show the same quantities, but now for a cubic lattice. It is easy to see that the results look qualitatively the same 
as for the 2d case. Furthermore, they reproduce the behavior obtained with 
other techniques such as QMC \cite{gubernatis1987spin} and NRG \cite{borda2007kondo}. Notice that the size of the chain 
corresponds to the radial dimension of the $d$-dimensional lattice. This means that a chain of length $N=100$ describes 
a square or a cube with a half-diagonal of the same size\cite{correlations}. 

\begin{figure}
\epsfxsize=8.50cm \centerline{\epsfbox{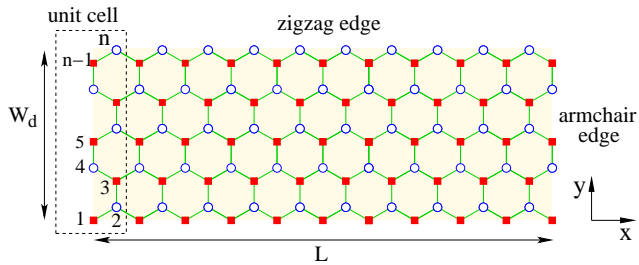}}
\caption{{\it Representation of a graphene nanoribbon.} We show a nanoribbon containing zigzag edges along its length $L$ and armchair edges along its width $W_d$. 
The two triangular sublattices are represented by squares and open circles. The unit cell in the armchair edge direction, containing $n$ sites, is circumscribed by a dashed box. 
By definition, a nanoribbon takes its classification (zigzag or armchair) from the geometry of its length edges.}
\label{figure8}
\end{figure}  

We next study the Anderson model Eq.~(\ref{Anderson_model}) for an adatom at the surface of a CNT. All simulations were 
done at half-filling. 
Unless stated otherwise, the parameters, in units of the hopping $t$, are $U=0.5$, $t'=\sqrt{0.05}$, and we work in the particle-hole symmetric point,{\it i.e.}, $V_g=-U/2$. 





 Figs.~\ref{figure7}(b) and (c), 
show the spin-spin correlations between the impurity and the first orbital $\langle S^z_{\rm d}~S^z_{\rm r_0} \rangle$, 
where ${\rm r_0}$ is the position of the adatom on the the CNT.
Results for a $\left(0,4\right)$ insulating CNT are shown in panel (b).
In this case, since the system is gaped and the correlation length is small, we achieve convergence for $N \sim 40$ ($1/N \sim 0.025$).  
The metallic case, for a $\left(4,4\right)$ CNT 
displays correlations that scale linearly in $1/N$. {These behaviors will be more thoroughly analyzed in the next section, in the context of graphene nanoribbons.}

\section{Graphene nanoribbons: Localized vs metallic states}


Since its discovery, graphene\cite{graphene1,graphene1b} (a monolayer of graphite with a honeycomb arrangement of carbon atoms) 
has become the subject of intense research due to its unusual electronic properties, in particular because of its massless 
Dirac spectrum\cite{graphene_review1,graphene_review2}. This makes graphene an ideal experimental system to look for exotic 
new physics, as well as a promising basis for novel nanoelectronics, raising the expectations for a post-silicon era \cite{berger2006,avouris2007,geim2007,lin2009}. 
Adding magnetic atoms or defects opens the possibility for spintronic applications, where not only the charge, but also 
the spin of the electrons can be manipulated in spintronic devices \cite{wolf2001,spintronics}.

The physics of diluted magnetic impurities in graphene is rich and constitutes an entire subject of research on its own right \cite{castro2009adatoms,Lars, shytov2009long}.
Isolated magnetic adatoms placed on graphene sheets have been studied experimentally as well as theoretically \cite{graphene_review1,graphene_review2}, 
and the properties of the Kondo ground state in graphene have been a subject of controversy. Experimental evidence of Kondo effect due to magnetic 
adatoms, such as cobalt, on top of graphene has been reported. Depending on the position of the adatom, different behaviors can be observed. 
For adatoms directly on top of carbon sites, a Fermi liquid behavior consistent with an $SU(2)$ Kondo effect has 
been {predicted and found to be in agreement with experimental results \cite{graphene_Kondo,cornaglia2009,sengupta2008,Jacob10}.} 
However, for adatoms at the center of an hexagon, the results are contradictory.
On one hand, based on symmetry arguments and DFT calculations, an $SU(4)$ Kondo effect was 
predicted \cite{wehling2010,kharitonov2013kondo}. On the other hand, renormalization group arguments show 
a two-channel Kondo effect with a characteristic non-Fermi liquid behavior \cite{cornaglia2009,schneider2011}. 
Moreover, the Kondo state does not depend only on the position of the adatom, but also on the band filling. 
By gating graphene, one can move the Fermi energy away from the Dirac point, to a region of the band with a linear density of states.


In this work, we use the nanoribbons just as a proof of concept, and not as an object of study, and we illustrate how the method can be used to identify and study localized edge states.

\subsection{Electronic properties of Graphene ribbons}
Graphene is an arrangement of carbon atoms in a 2d honeycomb structure, that can be described by two interpenetrating 
triangular lattices. This is shown in Fig.~\ref{figure8}, where a graphene nanoribbon with zigzag 
edges along its length, and armchair edges along its width, is represented. The two sub lattices are presented by squares and open circles. 

\begin{figure}
\vspace{0.2cm}
\epsfxsize=7cm \centerline{\epsfbox{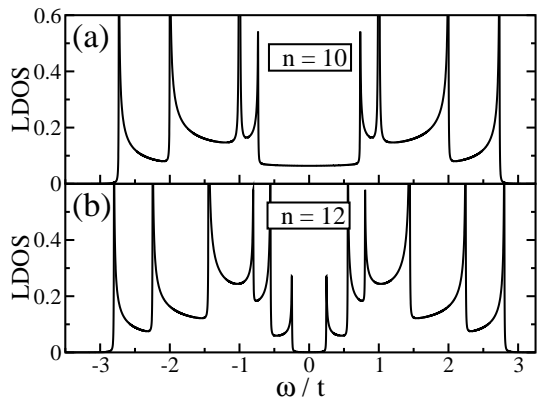}}
\caption{{\it Local density of states for armchair ribbons}. LDOS calculated along the $y$-direction of the ribbon 
(see text for details): (a) metallic, $n=10$, and (b) insulating, $n=12$ nanoribbons.}
\label{figure9}
\end{figure}  
 
It is well known that graphene nanoribbons and carbon nanotubes display similar characteristics. In the case 
of ribbons, their properties are determined by the geometric type (zigzag or armchair) of their 
edges (along their lengths) and by their width \cite{Dresselhaus}. Ribbons with 
zigzag edges are always metallic, albeit with localized edge-states, while armchair ribbons may be insulating or metallic, depending on their width.
Note that we classify nanoribbons by the shape of their longest side (zigzag or armchair) and by the number of sites $n$ in the unit cell \cite{note-cell}, as shown in Fig.~\ref{figure8}.
For the moment, we ignore correlation effects, and magnetism, focusing instead on the pure electronic properties. 
To reconstruct the band structure, we considered a tight-binding Hamiltonian with 7000 unit cells. 
Our unit of energy is the hopping $t\sim 2.8$eV and for simplicity we neglect the second-neighbors hopping $t'\sim 0.1t$ \cite{Lars}.
We chose as seeds for the Lanczos process, either atomic sites at the edge or at the center of the 
ribbon.
%

\begin{figure}
\epsfxsize=7cm \centerline{\epsfbox{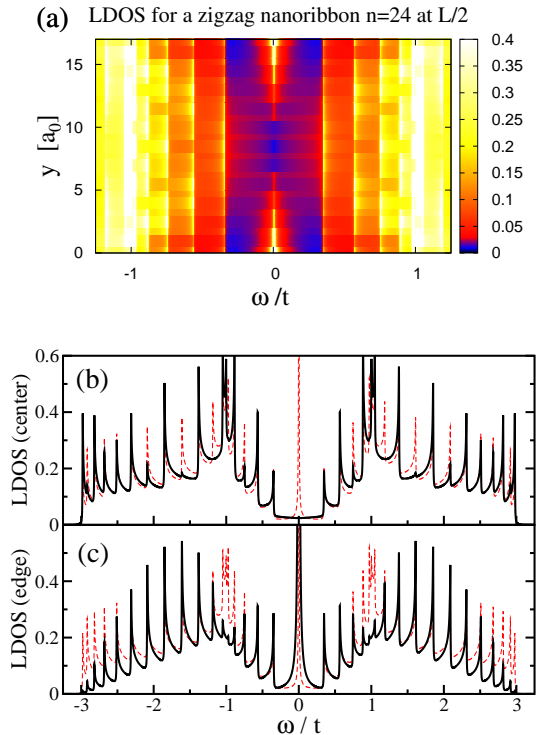}}

\vspace{0.2cm}
\epsfxsize=7cm \centerline{\epsfbox{figure10b.eps}}
\caption{{\it Local density of states for zigzag ribbons.} Panel (a) shows a color contour plot of 
the LDOS as a function of $\omega$ (horizontal axis, around the Fermi energy $\omega=0$) and along the $y$-direction 
of a zigzag ribbon with an $n=24$ unit cell (see Fig.~\ref{figure8} for the $x$ and $y$ axes directions). 
Details of the LDOS are shown at the center (b) and the edge (c) of 
the nanoribbon, respectively. In both panels, the total DOS, averaged over all sites {within the unit cell}, 
is shown as a (red) dashed line. A resonance at $\omega=0$ for the LDOS at the edge is consequence of the localized state. 
The absence, or presence, of this resonance, is the main qualitative difference between the (black) solid curves in panels (b) and (c).
We can see, in panel (a), how this resonance vanishes when we approach, from one of the edges, the center of the nanoribbon.}
\label{figure10}
\end{figure}  

In Fig.~\ref{figure9}(a), we show the total LDOS for two different cases of armchair nanoribbons. 
Note that the results shown are the average of the LDOS calculated at each site of the ribbon's unit cell. Therefore, it contains information about 
both the edge- and bulk-states simultaneously. The LDOS at each site of the unit cell was calculated after {performing 
a mapping which had as seed the site in question.}
In Fig.~\ref{figure10}, the LDOS for two different armchair nanoribbons is displayed: panel (a) corresponds to a metallic nanoribbon with $n=10$ 
sites in the unit cell, and panel (b) is for an insulating nanoribbon (having a gap $\sim 0.5t$) with $n=12$. 
As in the two examples for CNTs presented in the previous section, the LDOS for the nanoribbons shows a rich structure, involving Van Hove singularities. 
Increasing the number of sites in the unit cell increases the number of bands and hence the number of Van Hove singularities. 

Figure~\ref{figure10} shows the LDOS calculated for a zigzag nanoribbon with an $n=24$ unit cell. 
In panel (a), we show a color density plot of the LDOS as a function of $\omega$ (horizontal axis, around the Fermi energy $\omega=0$, at half-filling) 
and the $y$ coordinate along the width of the nanoribbon (vertical axis). Panels (b) and (c) present details of the LDOS at the edge ($y=0$) and at the center ($y=8.75~a_0$) of 
the nanoribbon, respectively. In panel (c), in the (black) solid curve, we can see a sharp peak at $\omega=0$ 
associated to the localized state at the edge. This peak decreases as we move towards the center of the nanoribbon 
[see panel (a)], and completely vanishes, as shown by the (black) solid curve in panel (b). 
Note that at, and around, $\omega=0$, a finite flat LDOS is observed, indicating that the nanoribbon 
presents a metallic region at its center. Furthermore, in panels (b) and (c), a (red) dashed line 
displays the total LDOS, averaged over all sites of the unit cell.

If we now couple a magnetic adatom to any of these systems, the behavior of the spin correlations will 
be determined by the electronic properties of the specific ribbon, and by the position of the impurity relative to the 
edges. Following Ref.~\onlinecite{Jacob10}, the adatom is placed on top of a carbon site, as shown in Fig.~\ref{figure2}(a).

\subsection{Results for armchair ribbons}
\begin{figure}
 \includegraphics[height=7.5cm, angle=-90]{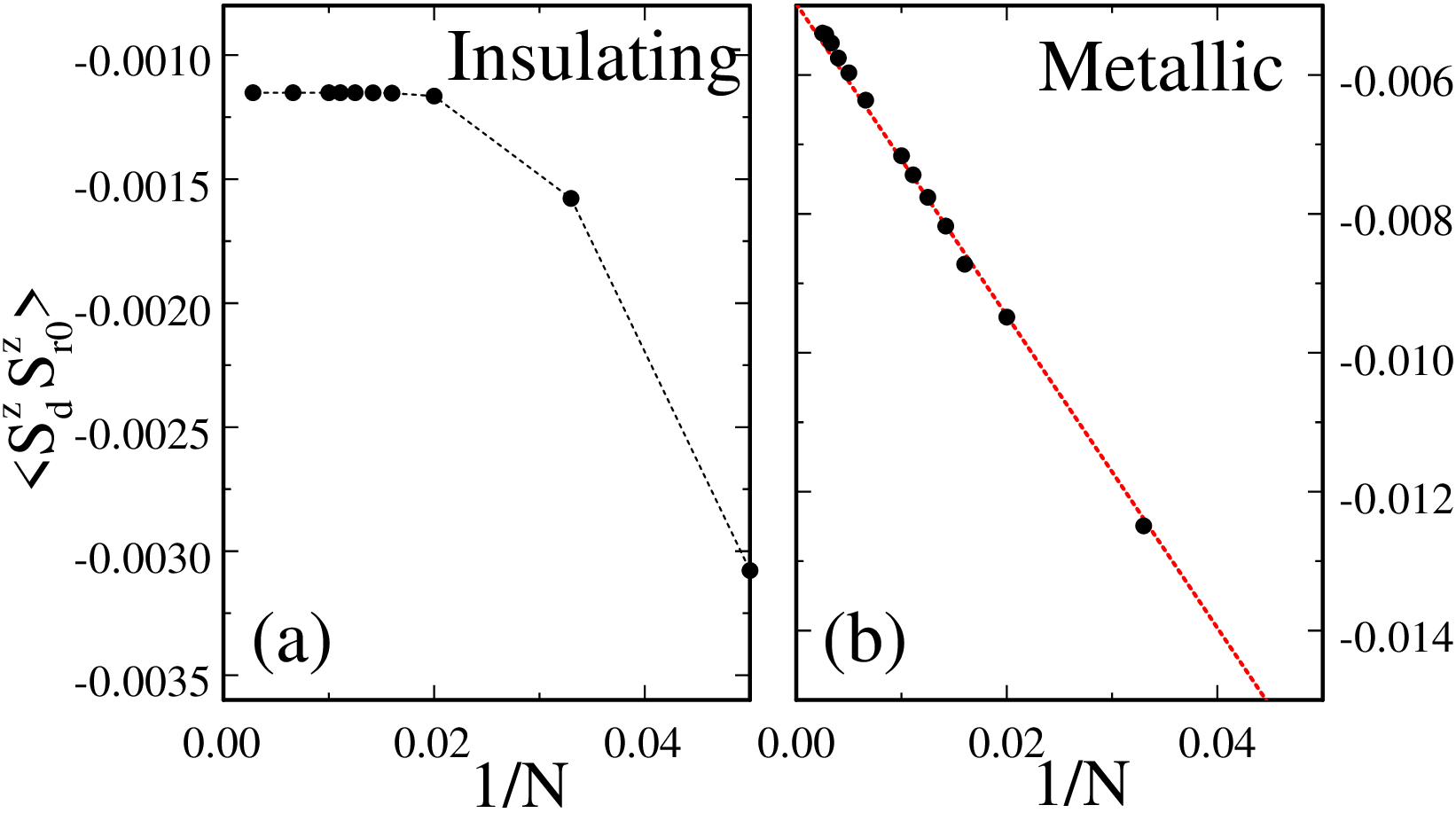}
 \includegraphics[height=7.5cm, angle=-90]{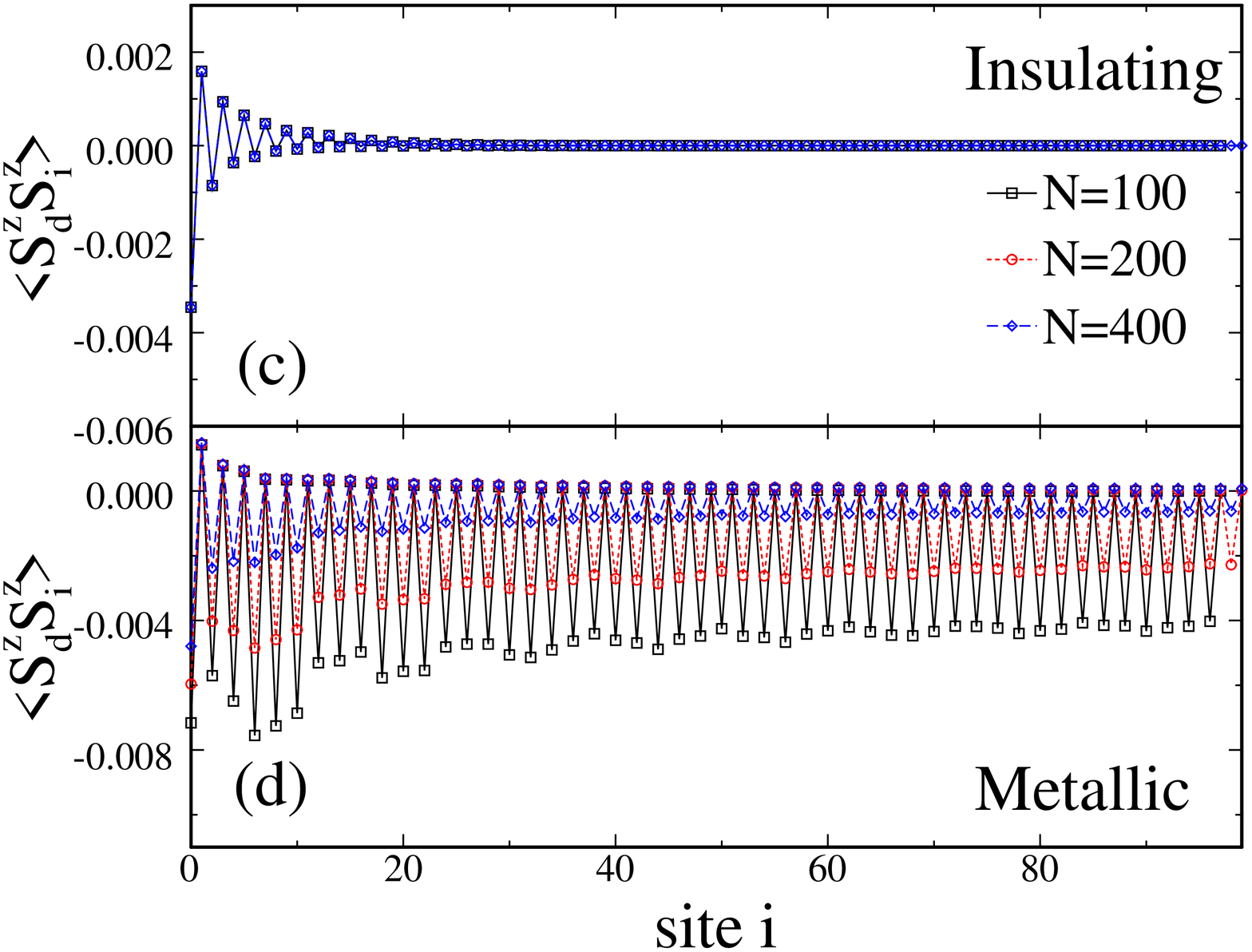}
\caption{{\it Spin correlations for an adatom in an armchair nanoribbon.} Panels (a) and (b) show the spin correlations 
between an impurity at the center of the nanoribbon and the first site of the chain as a function of the inverse of the system 
size for an insulating ($n=12$) and a metallic ($n=10$) nanoribbon, respectively. In the insulating case, 
the correlations saturate at $N \approx 50$, while in the metallic nanoribbon, the correlations 
scale linearly with $1/N$ (the extrapolated curve is shown as a (red) dashed line). In panels (c) and (d), the spin correlations between 
the impurity and the first $100$ Lanczos orbitals are shown. In panel (c) all cureves are indistinguishable. 
Even and odd sites along the chain represent orbitals in different sublattices of the original problem (see text for details).}
\label{figure11}
\end{figure}  

We first study the spin correlations for an adatom in an armchair nanoribbon. As in Sec.~\ref{SecIV}, we use the 
impurity Anderson model (IAM) to describe the problem. The total Hamiltonian is given by Eqs.~(\ref{hamiltonian}),~(\ref{Anderson_model}),~and~(\ref{hybridization}). 
We use the same parameters as before, $U=0.5$, $t'=\sqrt{0.05}$, and we work in the particle-hole symmetric point, {\it i.e.}, $V_g=-U/2$. 

In Figure~\ref{figure11}, we present results for a single Anderson impurity attached to insulating ($n=12$) and metallic ($n=10$) armchair nanoribbons. 
In both cases, we place the impurity at the center of the lattice.
Panels (a) and (b) show the spin-spin correlations between the impurity and the first site of the chain 
(after mapping), which corresponds the orbital is attached to the impurity in the original problem. 
For the insulating armchair nanoribbon, shown in panel (a), we observe a saturation of the spin correlations for system sizes larger than $N=50$ sites, 
which gives an indication of the correlation length in the system. As shown in panel (b), no such saturation occurs for a metallic nanoribbon. 
These results are very similar to the ones obtained in Fig.~\ref{figure7}. 

Panels (c) and (d) show the spin correlations between the impurity and the sites of the mapped chain for the insulating 
and metallic nanoribbon, respectively. In both cases, we observe alternating ferro and antiferromagnetic correlations. 
We recall that each site of the mapped chain is associated to an specific sublattice of the real system: the odd sites of the Lanczos chain correspond to the same sublattice as that of the site connected to the impurity. 
In panel (c), we again observe a fast convergence of the correlations with the system size, in the sense that 
all three curves (for $N=100$,$~200$, and $400$) are already indistinguishable. This is in agreement 
with the results in panel (a), where the saturation has already occurred for $N \approx 50$. 
These correlations decay virtually to zero at distances of 
the order of $50$ sites from the impurity, in agreement with the previous observation. 

In the metallic case, shown in panel (d), the ferromagnetic correlations decay fast as in the previous case. 
However, the correlations within the same sublattice as the impurity present a very slow decay. The correlation length is usually associated with the size of the Kondo cloud, which is of the order of $1/T_K$ \cite{Busser10}. 
However, the Kondo physics on its own cannot explain the strong finite-size dependence of the correlation length, which {\it decreases} as the system size {\it increases}. This seems to be a behavior dominated by the geometry of the problem. Since we are working with finite systems, this situation corresponds to that of a Kondo box, and indicates that we are far from the universal Kondo regime. This physics undoubtedly deserves further investigation.
%
Note that we also performed calculations for an impurity attached to the edge of an armchair nanoribbon (not shown here), essentially reproducing the same behavior. 

\subsection{Results for zigzag ribbons}
\begin{figure}
 \includegraphics[height=7.5cm, angle=-90]{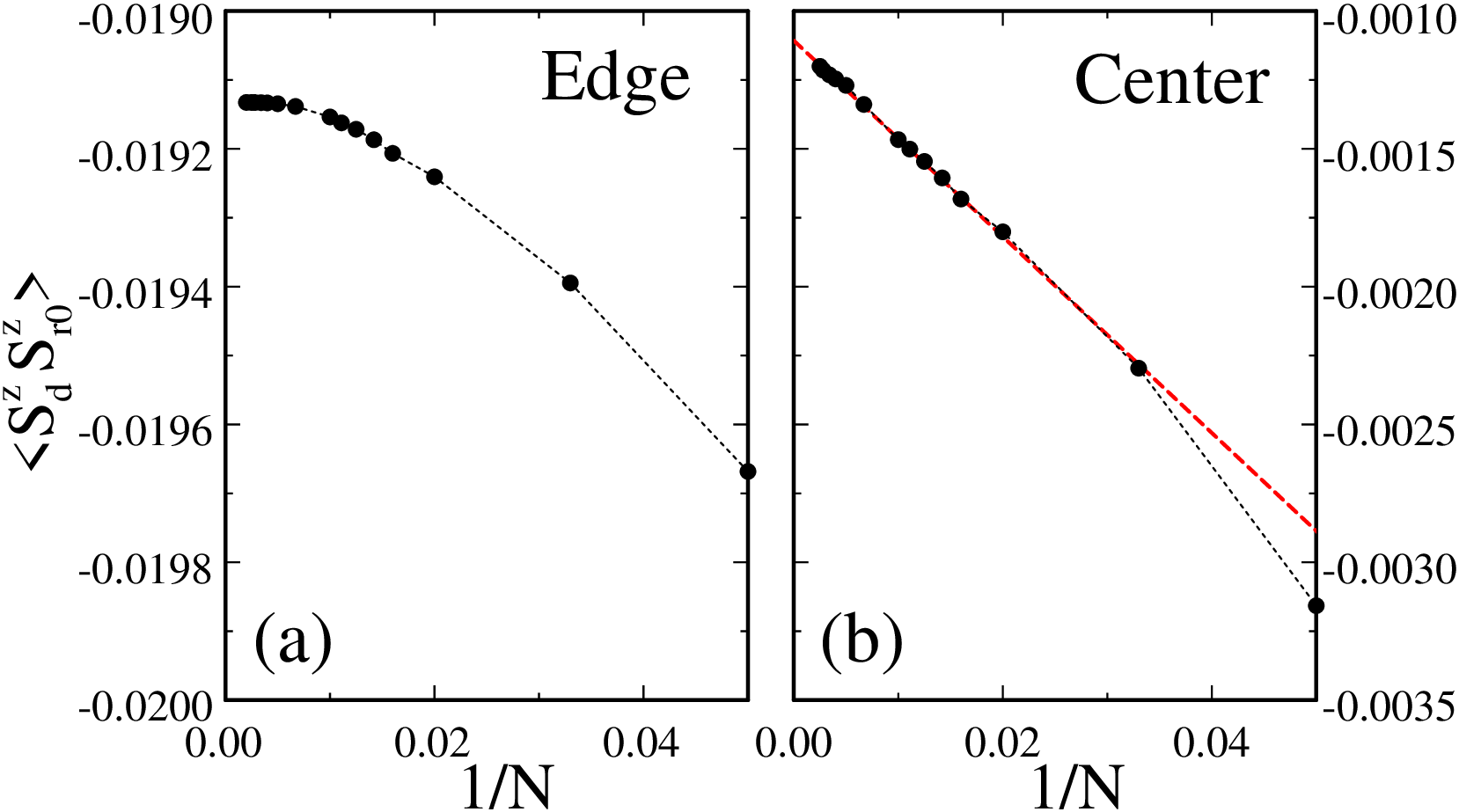}
 \includegraphics[height=7.5cm, angle=-90]{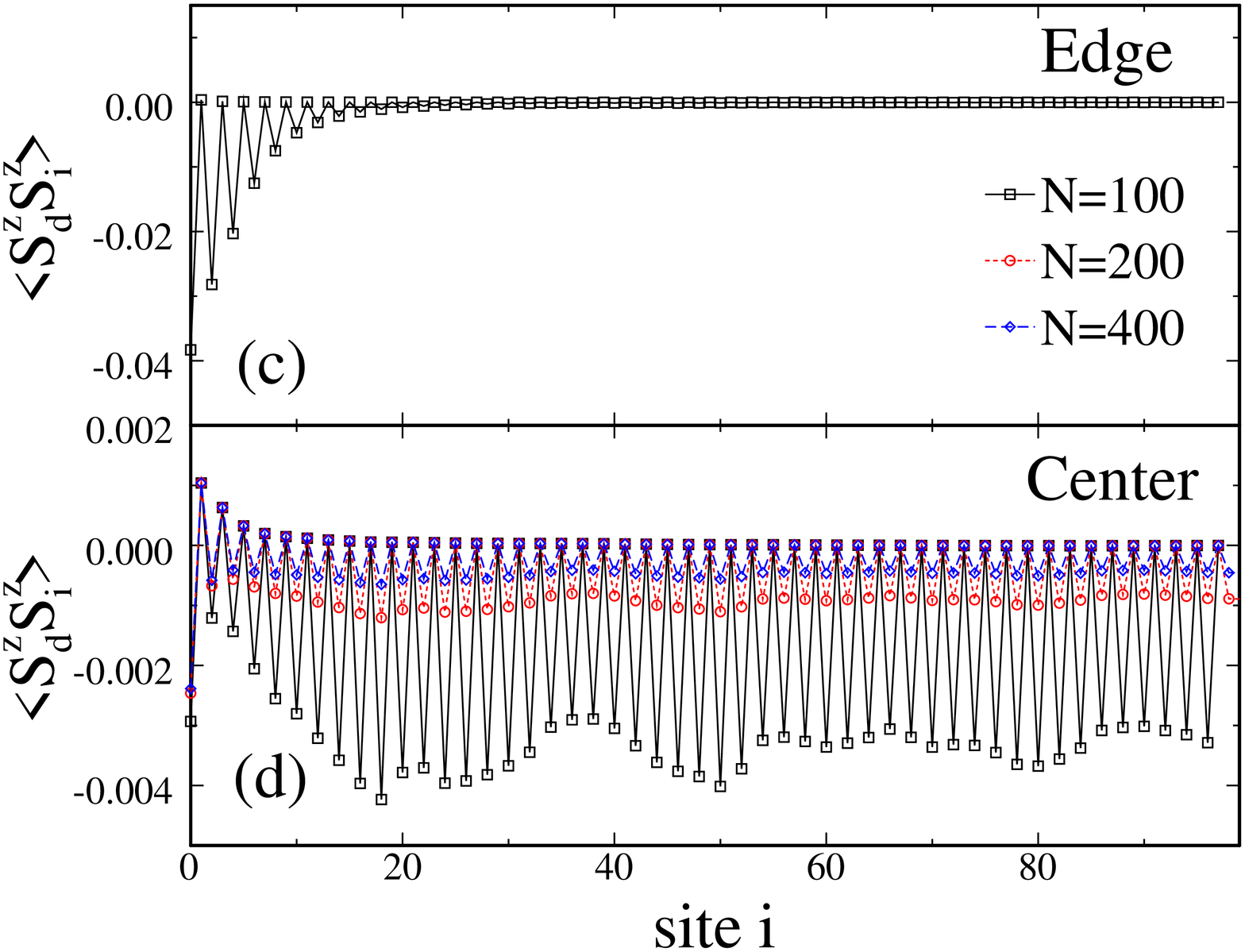}
\caption{{\it Spin correlations for an adatom in a zigzag nanoribbon.} Panels (a) and (b) present the spin correlations 
between the impurity and the first site of the Lanczos chain as a function of $1/N$. Panel (a) has results for 
an impurity coupled to an edge site and panel (b) for a central site. The edge states in zigzag nanoribbons 
are localized and correlations in (a) tend to saturate. In (b), since the impurity is at the center of the nanoribbon where the bulk states are metallic, no saturation occurs. Rather, one sees a linear dependence with 
$1/N$, as in the case of a metallic armchair nanoribbon [Fig.~\ref{figure11}(b)]. 
The correlations between the adatom as a function of the distance along the chain are 
shown in panels (c) for the impurity at the edge, and (d) for the impurity at the center (see text for details). 
}
\label{figure12}
\end{figure}  

We proceed in the same way to analyze the case of zigzag nanoribbons, for the same model and parameters as in 
the previous section.
In this instance, however, we need to differentiate the edge states from the bulk states of the nanoribbon. 
As shown in Fig.~\ref{figure10}, edge states are localized and bulk states are metallic. 
We study a system with an $n=24$ unit cell, placing the impurity at the edge for 
one simulation, and at the center for another.

Results are shown in Fig.~\ref{figure12}. In panels (a) and (b), we show the extrapolation to the 
thermodynamic limit for the spin correlation between the impurity and the first site of the chain, 
for an impurity located at the edge and at the center, respectively. 
For the case of the impurity attached to the edge of the zigzag nanoribbon, panel (a), we observe saturation at $N\sim100$. 
{The saturation is less well characterized than in the case of armchair insulating nanoribbons, likely due to  the proximity of metallic bulk states.}
When the impurity is coupled to a site in the center of the nanoribbon, as shown in panel (b), 
the correlations scale linearly in $1/N$, as expected for the gapless case.  

Panels (c) and (d) show the spin-spin correlations between the impurity and the Lanczos orbitals of the chain, 
as a function of the distance from the impurity. 
The results for the impurity at the center are plotted in (d), showing similar behavior as the metallic armchair ribbon. 
Remarkably, the correlations do not display a noticeable decay in the observed range. However, the finite-size effects are quite dramatic, as seen in the results for larger systems. As in the previous section, we attribute this behavior to the fact that we are in a Kondo box, far from the universal Kondo regime.

\section{Conclusions}

In this work we propose a canonical transformation based on the Lanczos method to map complex single particle band structures, onto equivalent 1d systems. 
We point out that even though the recursion method has been available for some time, its application has typically been limited to spectral properties of non-interacting systems \cite{roche1999quantum}.
We introduce many-body interactions in the Hamiltonian that require the use of appropriate techniques to solve the equivalent problem, such as DMRG, ECA, or QMC. For the case of single impurities, this transformation preserves the form of the hybridization Hamiltonian between the 
impurity and the lattice, and gives us the resolution to study the behavior of correlations in real space.

We have shown that complex systems, like CNTs, either metallic or with insulating gaps, or presenting van Hove singularities, 
can be perfectly treated with this method. Using DMRG, we are able to study fairly large systems. 
As a benchmark and illustration, we show that large $d$-dimensional systems can be studied with little effort, once their band structure is mapped onto the effective 1d Hamiltonian. 

In addition, we have studied adatoms on carbon nanotubes and graphene nanoribbons. The complex and rich electronic properties 
of these systems, as well as their geometry, make them especially suitable for the application of our method.
We can clearly distinguish between insulating and metallic regimes, or edge and bulk states, 
depending on the geometry of the system, and the position of the impurity.

It is important to point out that extrapolations to the thermodynamic limit cannot be taken trivially.
On the one hand, when the system under study is insulating, correlations decay exponentially, and we can easily obtain thermodynamic results. 
 However, this is not as simple for the case of 
impurity problems in metallic hosts. We work with finite systems, far from the universal Kondo regime.
This regime, when $N\to \infty$, can only be reached with NRG, a method
that is precisely designed to study this limit.
 Despite this, the short distance behavior can be obtained fairly accurately using our approach.

When treating finite lattices, we need to recall that we are 
working in a Kondo-box, where the inter-level spacing of the discrete band structure 
introduces an additional energy scale into the problem. 
We point out that carbon nanotubes and graphene nanoribbons {\it are} finite and could be relatively small in size. Our method is ideal to capture the physics of these problems.

Using the Lanczos transformation on problems with many-body interactions and phonons can be done in a similar fashion\cite{Bonca1999,Bonca2007,Vidmar2011,Mierzejewski2011,Vidmar2011b}, but the resulting Hamiltonian will be quite complicated and non-local. An alternative is to introduce correlations at the dynamical mean field level, as proposed in Ref.\onlinecite{garcia2004dynamical} (See also Ref.\onlinecite{dargel2012lanczos}).

In principle, any quadratic Hamiltonian can be mapped onto an equivalent 1d system following our prescription. 
This opens the doors to problems with spin-orbit interaction, and superconductors (at the mean field/BdG level). 
Possible applications of the mapping include topological insulators, where the bulk and the boundary have very different electronic properties \cite{roche1999quantum,lherbier2008transport}. 

Even though we have used DMRG to solve the effective one-dimensional Hamiltonian, one could use alternative methods, such 
as ECA, or QMC, as the problem has no sign problem. One could also use these techniques to study 
time-dependent \cite{white2004real,Daley2004}, and thermodynamic properties \cite{feiguin2005finite}, or to calculate spectral functions \cite{barthel2009spectral, feiguin2010spectral}.

The method can also become a powerful tool to study quantum chemistry problems in which a magnetic atom is embedded into 
a system that may be described by a H\"uckel-like theory\cite{busser2012designing}. Moreover, one could use it in conjunction with ab-initio band 
structure calculations, incorporating the information about the bulk system into the effective 1d chain, and the 
hybridization terms in the structure of the seed state, paving the way toward realistic first principles modeling and 
a better understanding of correlation effects in quantum impurity problems.

\begin{acknowledgments}
We thank K. Al-Hassanieh, F. Heidrich Meisner, and L. Vidmar for helpful discussions.
CAB was supported by the {\it Deutsche Forschungsgemeinschaft} (DFG) through FOR 912 under grant-no. HE5242/2-2. 
GBM acknowledges financial support from the NSF under Grants DMR-0710529, DMR-1107994, and MRI-0922811. 
GBM also thanks the hospitality of the Institut f\"ur Theorie der Statistischen Physik, RWTH Aachen University, Aachen, Germany, 
where part of this work was performed under a DAAD fellowship. AEF acknowledges NSF support through grant DMR-1339564.
\end{acknowledgments}


\begin{thebibliography}{97}
\bibitem[{cor()}]{correlations}
\bibinfo{note}{Note: A detailed study of the correlations will be presented
  elsewhere}.
\bibitem[{not()}]{note-cell}
\bibinfo{note}{Note that, as we use OBC for all our lattices, when reference is
  made to a unit cell, we mean a unit cell at the center of the cluster, unless
  specified otherwise.}
\end{thebibliography}

\end{document}